\documentclass[aps,prb,twocolumn,superscriptaddress,longbibliography]{revtex4-1}

\usepackage{hyperref} 
\usepackage{graphicx}
\usepackage{amsmath}
\usepackage{amssymb}
\usepackage[dvipsnames,svgnames,x11names,hyperref]{xcolor}
\hypersetup{colorlinks=true,linkcolor=NavyBlue,urlcolor=NavyBlue,  citecolor=NavyBlue}
\DeclareGraphicsExtensions{.png,.jpg,.eps,.pdf}

\usepackage[caption=false]{subfig}

\usepackage[T1]{fontenc}

\renewcommand{\vec}[1]{\boldsymbol{\mathbf{#1}}}

\newcommand{\hc}{^{\dagger}}

\begin{document}

\title{Correlation-induced valley topology in buckled graphene superlattices}

\author{Antonio L. R. Manesco}
\email{am@antoniomanesco.org}
\affiliation{Computational Materials Science Group (ComputEEL),
Escola de Engenharia de Lorena, Universidade de São Paulo (EEL-USP),
Materials Engineering Department (Demar), Lorena – SP, Brazil
}
\affiliation{Kavli Institute of Nanoscience, Delft University of Technology, Delft 2600 GA, The Netherlands}

\author{Jose L. Lado}
\affiliation{Department of Applied Physics, Aalto University, Espoo, Finland}

\date{\today}

\begin{abstract}
    Flat bands emerging in buckled monolayer graphene superlattices have been recently shown to realize correlated states analogous to those observed
    in twisted graphene multilayers.
    Here, we demonstrate the emergence of valley topology driven by competing electronic correlations in buckled graphene superlattices. We show, both by means of atomistic models and a low-energy description,
    that the existence of long-range electronic correlations leads to a competition between antiferromagnetic and charge density wave instabilities, that can be controlled by means of screening engineering.
    Interestingly, we find that the emergent charge density wave has a topologically non-trivial electronic structure, leading to a coexistent quantum valley Hall insulating state. In a similar fashion,
    the antiferromagnetic phase realizes a spin-polarized quantum valley-Hall insulating state.
    Our results put forward buckled graphene superlattices as a new platform to realize interaction-induced
    topological matter.
\end{abstract}

\maketitle

\section{Introduction}
\label{sec:qvh-intro}

Mesoscopic systems provide a highly powerful platform  
to design quantum matter,\cite{zhangMoirQuantumChemistry2020, wuHubbardModelPhysics2018, chenTunableMoireSuperlattice2019, ladoDesignerQuantumMatter2021, Andrei2021} with the paradigmatic example of artificial
topological superconductivity.\cite{kitaevFaulttolerantQuantumComputation2003, lutchyn2010majorana, lawMajoranaFermionInduced2009, zhangLargeZerobiasPeaks2021, zhangNextStepsQuantum2019, mourikSignaturesMajoranaFermions2012, Kezilebieke2020,san-joseMajoranaZeroModes2015, oregHelicalLiquidsMajorana2010, fuSuperconductingProximityEffect2008}
Moire two-dimensional materials have risen as a tunable platform
to engineer states of matter,\cite{Andrei2021} ultimately allowing to explore a variety of controllable correlated states.\cite{songAllMagicAngles2019, xieTopologyBoundedSuperfluidWeight2020, daliaoCorrelationInducedInsulatingTopological2021, choiCorrelationdrivenTopologicalPhases2021}
This emergence of tunable correlations stems from the quench of kinetic energy in emergent flat bands, controllable by twist engineering\cite{kauppilaFlatbandSuperconductivityStrained2016, kopninHightemperatureSurfaceSuperconductivity2011, tangStraininducedPartiallyFlat2014}
A variety of twisted van der Waals materials have been demonstrated in this direction, including bilayers, trilayers and tetralayers\cite{ganiSuperconductivityTwistedGraphene2019, caoCorrelatedInsulatorBehaviour2018a,Shen2020,Liu2020,Park2021,2021arXiv210312083C}.

\begin{figure}[!t]
    \subfloat[\label{fig:qvh-scheme}]{\includegraphics[width=0.48\linewidth]{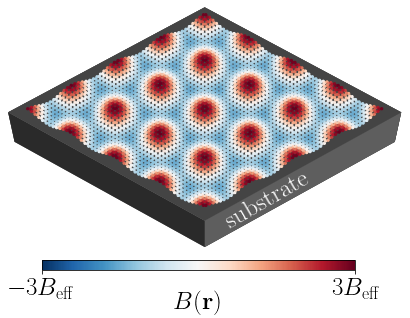}}
    \subfloat[\label{fig:ldos}]{\includegraphics[width=0.48\linewidth]{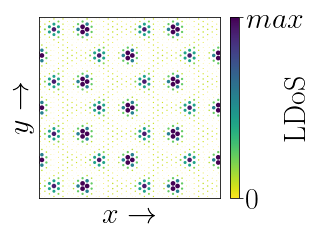}}\\ \vspace*{-0.5cm}
    \subfloat[\label{fig:full_free}]{\includegraphics[height=.175\textheight]{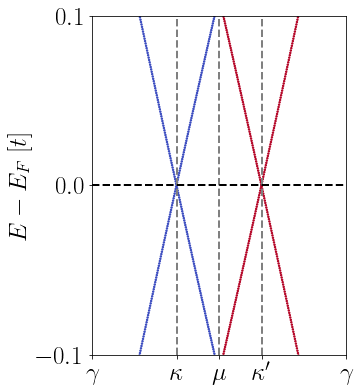}}
    \subfloat[\label{fig:full_lattice_valleys}]{\includegraphics[height=.175\textheight]{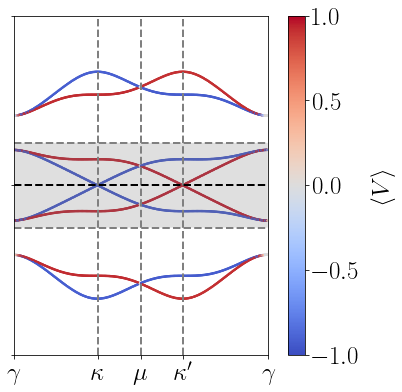}}
    \caption{
        (a) Scheme of the buckled graphene superlattice studied.
        The $B$-field is the same as in Eq. \ref{eq:B-field}.
        (b) Local density of states from full-lattice calculations.
        It is possible to note the emerging honeycomb structure.
        Valley-projected band structures from a full-lattice simulation of a buckled graphene superlattice (c) in the absence of strain
        and (d) in the buckled state.
        The dashed grey regions indicate the active bands for which we derive the effective model.
    }
    \label{fig:full_lattice}
\end{figure}

Beyond the wide family of twisted moire multilayer heterostructures\cite{Andrei2021}, monolayer graphene 
has also been experimentally shown to realize moire-induced
correlation physics in the single layer limit.\cite{maoEvidenceFlatBands2020}
The field of straintronics, \emph{i.e.}, the control of electronic properties of materials with strain,\cite{bukharaevStraintronicsNewTrend2018} has shown different methods to create two-dimensional periodically-strained superlattices, from substrate engineering\cite{jiangVisualizingStrainInducedPseudomagnetic2017} to inducing buckling transitions during fabrication.\cite{maoEvidenceFlatBands2020}
From a low-energy perspective, strain fields act as valley-dependent pseudo-magnetic fields,
leading to the emergence of pseudo-Landau levels.\cite{vozmedianoGaugeFieldsGraphene2010a, lowStrainInducedPseudomagneticField2010, RAMEZANIMASIR201376,Peltonen2020}
From a critical value of in-plane strain, the elastic energy is spontaneously reduced with out-of-plane distortions,\cite{baoControlledRippleTexturing2009, caiPeriodicPatternsEnergy2011, cerdaGeometryPhysicsWrinkling2003} as depicted in Fig. \ref{fig:qvh-scheme}.
The electronic structure reconstruction due to the strain field leads to the formation of nearly flat bands.\cite{maoEvidenceFlatBands2020, milovanovicPeriodicallyStrainedGraphene2019}
Moreover, the bandwidth suppression enhances the interaction effects and leads to electrically-controllable correlated phases.\cite{maoEvidenceFlatBands2020, manescoCorrelationsElasticLandau2020}

Here we show that buckled graphene superlattices yield correlation-induced
topological states, stemming from the combination of pseudo-Landau levels and
non-local electronic interactions.\cite{maoEvidenceFlatBands2020, manescoCorrelationsElasticLandau2020}
In particular, we show that the low energy states generated by the buckling
(Fig. \ref{fig:ldos}) shows an emergent low-energy honeycomb structure.
Also, similarly to free-standing graphene, the bandstructure (Fig. \ref{fig:full_lattice_valleys}) has Dirac cones at the corners of the mini-Brillouin zone.\cite{manescoCorrelationsElasticLandau2020}
We derive the low-energy model describing the bands closer to the Fermi energy (Sec. \ref{sec:qvh-system}), to explore the impact of electron-electron interactions, and show the existence of 
charge density wave and antiferromagnetic ground states (Sec. \ref{sec:qvh-interactions}).
Interestingly, these phases driven by electronic interactions
show finite valley Chern numbers, and associated topological surface states. 
We finally demonstrate the robustness of our model
by comparing it with full atomistic selfconsistent calculations,
showing analogous phenomenology
as the one predicted by the effective model.
Our results demonstrate that buckled graphene monolayer can sustain a rich family of correlated
topological states, realizing analogous physics to twisted graphene multilayers in the single monolayer limit.

\section{The system}
\label{sec:qvh-system}

\begin{figure}[!t]
  \subfloat[\label{fig:kane-mele}]{\includegraphics[height=.15\textheight]{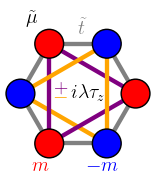}}
  \subfloat[\label{fig:effective_valleys}]{\includegraphics[height=.15\textheight]{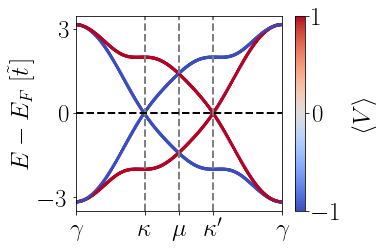}}
  \caption{
  (a) Schematic representation of the effective model in Eq. \ref{eq:topo-effetive_tb}.
  (b) Valley-projected bandstructure of the effective model described by the Hamiltonian \ref{eq:topo-effetive_tb} with $m=M=t$.}
  \label{fig:effective_bands}
\end{figure}

We first review the effective model for the buckled graphene superlattice, depicted in Fig. \ref{fig:qvh-scheme}.\cite{manescoCorrelationsElasticLandau2020, maoEvidenceFlatBands2020}
We take the graphene tight-binding Hamiltonian
\begin{align}
\label{eq:tb-ham}
\mathcal{H} = -t \sum_{\langle i, j \rangle} \sum_s \psi_{i s}^{\dagger}\psi_{j s} ,
\end{align}
where $t$ is the nearest-neighbor hopping constant, $i$ and $j$ denote different sites, $s$ denotes spin, $\langle i, j \rangle$ indicates the summation over nearest-neighbors, $\psi_{is}$ is the annihilation and $\psi_{is}\hc$ is the creation operator in position $i$ with spin $s$.

Under in-plane strain, the system undergoes a buckling transition, modifying the hoppings as\cite{manescoCorrelationsElasticLandau2020}
\begin{align}\label{eq:hoppings}
  \delta t_n = - \frac{\sqrt{3} e v_F L_M}{4 \pi} \sin(\vec{b}_n \cdot \vec{r}),
\end{align}
where $L_M$ is the superlattice size, and $v_F$ is the Fermi velocity. The three vectors
\begin{align}
  \vec{b}_1 &= \frac{2\pi}{L_M}\left(-\frac{1}{\sqrt{3}}, 1, 0 \right),\\
  \vec{b}_2 &= \frac{2\pi}{L_M}\left(\frac{2}{\sqrt{3}}, 0, 0 \right),\\
  \vec{b}_3 &= \frac{2\pi}{L_M}\left(-\frac{1}{\sqrt{3}}, -1, 0 \right)
\end{align}
point along the same direction of each hopping vector.

In the $k\cdot p$ approximation, the Hamiltonian with hoppings given by Eq. \ref{eq:hoppings} corresponds to a pseudo-magnetic field with the form\cite{maoEvidenceFlatBands2020}
\begin{align}
  \label{eq:B-field}
  B(\vec{r}) = B_{\text{eff}} \sum_{n=1}^3 \cos(\vec{b}_n\cdot\vec{r}).
\end{align}
Under zero strain, the electronic structure is folded in the mini-Brillouin zone defined by the $\vec{b}_n$ vectors (see Fig. \ref{fig:full_free}).
As the strain takes a finite value, avoided crossings are formed, creating mini-bands (see \ref{fig:full_lattice_valleys}) which we interpret as pseudo-Landau bands.\cite{maoEvidenceFlatBands2020, manescoCorrelationsElasticLandau2020}
Hence, quasiparticles feel a bandwidth quench.

We perform the valley projection in full-lattice calculations (Figs. \ref{fig:full_lattice_valleys} and \ref{fig:qvhi}) computing the expectation value of the modified Haldane coupling, $\langle V \rangle = \langle \Psi | V | \Psi \rangle$, with\cite{PhysRevLett.120.086603,PhysRevLett.121.146801,PhysRevResearch.2.033357}
\begin{align}
    V = \frac{i}{3 \sqrt{3}} \sum_{\langle \langle i, j \rangle \rangle} \eta_{ij} (\sigma_z)_{ij} \psi_i\hc \psi_j,
\end{align}
where $\eta_{ij} = \pm 1$ for clockwise/anticlockwise hopping, $\langle \langle i, j \rangle \rangle$ denotes a sum over second-neighbors, and $\sigma_z$ acts om sublattice degrees of freedom.

From the local density of states plot in Fig. \ref{fig:ldos}, obtained with full-lattice tight-binding calculations,\cite{manescoCorrelationsElasticLandau2020} it is possible to infer that the system has an emerging honeycomb superlattice.
The Wannier sites are localized at the minima and maxima of $B(\vec{r})$ since the characteristic length $\sqrt{\hbar / e B(\vec{r})}$ is smaller near the extrema.
The two extrema (minimum and maximum) correspond to the two sublattices of this effective honeycomb structure.
To reduce the computational cost of our numerical calculations, we now focus on the
low-energy model of these Wannier states.
We focus on the active bands closer to the Fermi energy, highlighted in Fig. \ref{fig:effective_valleys}.
Namely, we derive an effective model for the bands within the $[-0.025t, 0.025t]$ energy window in Fig. \ref{fig:full_lattice_valleys}.
This approach is analogous to low-energy models of twisted-bilayer graphene.

From both the space-dependent hopping constants (Eq. \ref{eq:hoppings}) and density of states (Fig. \ref{fig:ldos}), we conclude that the system is invariant under $C_3$-rotations.
Moreover, the bandstructure in Fig. \ref{fig:full_lattice_valleys} suggest that valley number is a conserved quantity.
Finally, in the absence of electronic interactions, the system has time-reversal symmetry.
With the current constraints, 
we find that the family of honeycomb Hamiltonians restricted to these symmetries is\cite{varjasQsymmAlgorithmicSymmetry2018}
\begin{align}\label{eq:topo-effetive_tb}
  \mathcal{H} &= - \mu \sum_{i} \sum_{s, \tau} c_{is\tau}\hc c_{is\tau}
  + m \sum_i\sum_{s, \tau} (\sigma_z)_{ii} c_{is\tau}\hc c_{is\tau}\\
  &-\tilde{t} \sum_{s, \tau} \sum_{\langle i, j \rangle} c_{is\tau}\hc c_{js\tau}
  + i \lambda \sum_{s, \tau} \sum_{\langle \langle i, j \rangle \rangle} (\tau_z)_{\kappa \kappa} \eta_{ij} c_{is\tau}\hc c_{js\tau} \nonumber 
\end{align}
where $c_{is\tau}\hc$ are creation and $c_{is\tau}$ annihilation operators operators at the site $i$, sublattice $\sigma$, valley $\tau$, and spin $s$.
The Pauli matrices $\sigma_i$ and $\tau_i$ act on sublattice and valley degrees of freedom.
The onsite energy and the hopping constants are denoted by $\tilde{\mu}$ and $\tilde{t}$ to distinguish to the atomistic model.
There is also a sublattice imbalance $m$ and a valley-dependent second-neighbors hopping $\lambda$.
An scheme of this model is shown in Fig. \ref{fig:kane-mele}.
Note that, since the Brillouin zone of this system corresponds to the mini-Brillouin zone from the atomistic model, there is an extra \emph{mini-valley} degree of freedom corresponding to the two nonequivalent points $\kappa$ and $\kappa'$ in effective model Brillouin zone.

It is visible that the Hamiltonian of Eq. \ref{eq:topo-effetive_tb} is equivalent to the Kane-Mele (KM) model: it consists on the tight-binding model of a honeycomb structure with a sublattice imbalance and a second-neighbors hopping that depends on the valley isospin.
The mapping between both models is made by identifying the valley isospin in the buckled superlattice to spin in KM model (spin$_{KM}$ $\to$ valley$_{\text{buckled}}$), as well as identifying the mini-valleys $\kappa$ and $\kappa'$ in the buckled system to the valleys $K$ and $K'$ in KM model (valley$_{KM}$ $\to$ mini-valley$_{\text{buckled}}$).\cite{kaneQuantumSpinHall2005, kaneTopologicalOrderQuantum2005}
As shown in Fig. \ref{fig:effective_valleys}, the energy dispersion is similar to the bandstructure of the full system [Fig. \ref{fig:full_lattice_valleys}] when $m = 3\sqrt{3} \lambda =: M$.
Therefore, the non-interacting strained system (without symmetry breakings) is enforced to have $M=m=\tilde{t}$ due to its gapless nature.
Note that arbitrarily small variations of $m / M$ open a gap in the effective model.
If $\delta(m/M) > 0$, the system becomes a trivial insulator.
On the other hand, $\delta(m/M) < 0$ opens a topological gap and the system becomes a quantum valley Hall insulator, in analogy to the spin Hall insulator phase in the KM model.

\begin{figure}[t!]
    \includegraphics[width=0.9\linewidth]{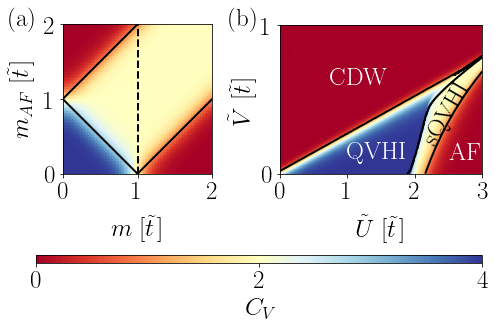}
  \caption{
  Valley Chern number dependence on (a) sublattice imbalance $m$ and antiferromagnetic mass $m_{\mathrm{AF}}$
  for the effective model, taking constant $\lambda$.
  Panel (b) shows the valley Chern number obtained after including interactions
  in the interactiong model, as a function of the Hubbard constant $\tilde{U}$ and nearest-neighbors interactions $\tilde{V}$.
  Solid lines indicate a topological phase transition and dashed lines indicate $m=M$, which
  for $m_{\mathrm{AF}}$ corresponds to the non-interacting strained system.
  }
  \label{fig:topo}
\end{figure}

Since the topographic shape of the buckling has the same functional form of $B(\vec{r})$, out-of-plane displacement fields lead to a modulation of onsite energies as:\cite{manescoCorrelationsElasticLandau2020}
\begin{align}
  \mu(\vec{r}) = \mu_0 \sum_{n=1}^3 \cos(\vec{b}_n\cdot\vec{r}).
\end{align}
Thus, near the maxima of $B(\vec{r})$, $\mu(\vec{r}) \approx 3 \mu_0$, and $\mu(\vec{r}) \approx 3 \mu_0 / 2$ near the minima of $B(\vec{r})$.
From the effective model perspective, the onsite energy modulation is:
\begin{align}\label{eq:displacement_effective}
  \mathcal{H}_{\mathrm{elec}} = 3 \mu_{\mathrm{elec}} \sum_{i \in A}\sum_{s. \tau} c_{is\tau}\hc c_{is\tau} - \frac{3 \mu_{\mathrm{elec}}}{2} \sum_{i \in B}\sum_{s, \tau} c_{is\tau}\hc c_{is\tau},
\end{align}
where $\mu_{\mathrm{elec}} \propto \mu_0$.
The sum over $i$ is performed on different sublattices in Eq. \ref{eq:displacement_effective}, since the corresponding Wannier sites are located at the maxima and minima of $\mu(\vec{r})$.
This extra term modifies the Hamiltonian as $m \to m + 3 \mu_{\mathrm{elec}} / 2$, and $\tilde{\mu} \to \tilde{\mu} - 3 \mu_{\mathrm{elec}} / 2$.
Therefore, out-of-plane displacement fields might be used as a knob to control the ratio $m/M$, ultimately working as an electric control of the system's topology, as shown in Fig. \ref{fig:topo}.

\begin{figure}[t!]
    \subfloat[\label{fig:af}]{\includegraphics[width=0.3\linewidth]{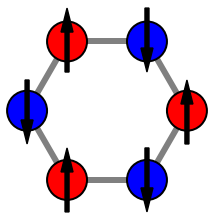}}
    \hspace{0.1\linewidth}
    \subfloat[\label{fig:cdw}]{\includegraphics[width=0.3\linewidth]{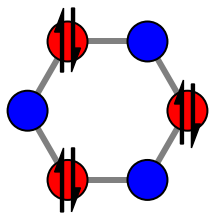}}\\ \vspace*{-0.5
    cm}
    \subfloat[\label{fig:qvh-phase_diagram}]{\includegraphics[height=0.175\textheight]{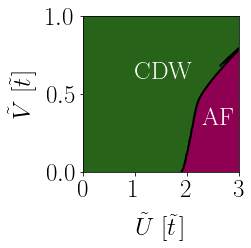}}
    \subfloat[\label{fig:qvh-gap}]{\includegraphics[height=0.175\textheight]{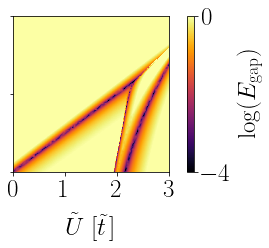}}
  \caption{
  Illustration of (a) antiferromagnetic and (b) charge density wave groundstates.
  (c) Phase diagram as a function of the coupling constants $\tilde{U}$ and $\tilde{V}$.
  The groundstate is a charge density wave (CDW) unless $\tilde{U}$ is sufficiently high.
    For $\tilde{U}$ higher than a critical value, an antiferromagnetic (AF) order develops.
    (d) System gap as a function of the electron-electron couplings.
    It is visible that the gap closes outside the region in which the broken symmetry changes.
  }
  \label{fig:qvh-phases}
\end{figure}

\section{Interaction-driven quantum valley Hall effect}
\label{sec:qvh-interactions}

The reduced bandwidth reduction due to the pseudo-magnetic field has been shown to lead to a correlated phase.\cite{maoEvidenceFlatBands2020,manescoCorrelationsElasticLandau2020}
Yet, due to the degeneracy of
the low energy states,
different groundstates may be realized in the system,
depending on the range and strength of interactions.\cite{kharitonovPhaseDiagramEnsuremath2012a, goerbigElectronicPropertiesGraphene2011a, youngTunableSymmetryBreaking2014}
The computational cost of full-lattice calculations makes an extensive investigation of possible groundstates impractical.
Hence, the reduced computational cost with an effective model allows us to explore the phase diagram as a function of electronic interactions.

To investigate the phase diagram of buckled graphene, we now include electronic interactions in the low energy model
\begin{align}
    \mathcal{H}_{\mathrm{int}} = \tilde{U} \sum_{\substack{\alpha, \beta \\ \alpha \neq \beta}} \sum_{i} n_{i \alpha} n_{i \beta} + \tilde{V} \sum_{\langle i, j \rangle} \sum_{\alpha, \beta} n_{i \alpha} n_{j \beta}
\end{align}
where $\tilde{U}$ is the onsite Hubbard interaction, $\tilde{V}$ is the nearest-neighbor interaction, $n_{i \alpha}:=c_{i \alpha}\hc c_{i \alpha}$ is the number operator at the Wannier site $i$.
The subindices $\alpha$ and $\beta$ are a short-hand notation to include both valley and spin degrees of freedom.

We solve this Hamiltonian by means of a mean-field approximation. Namely, we make
\begin{align}
    \mathcal{H}_{\mathrm{int}} \approx \mathcal{H}_{\mathrm{MF}} = \sum_{i, j, \alpha, \beta} \chi_{ij\alpha\beta} c_{i \alpha}\hc c_{j \beta},
\end{align}
and find $\chi_{ij\alpha\beta}$ self-consistently.
First, it is important to note that the interaction strengths $\tilde{U}$ and $\tilde{V}$ depend on the screening
created by the substrate of the buckled structure,\cite{Stepanov2020} and as such can be controlled
by screening engineering.\cite{Stepanov2020,PhysRevB.100.161102,PhysRevResearch.3.013265,PhysRevB.102.115111}
In the following, we will explore the potential symmetry broken states as a function of the
two interaction strengths, keeping in mind that such values would be controlled by substrate
engineering.
As we change the ratio of the local and non-local interaction, we see that there are two different groundstates, shown in the phase diagram of Fig.~\subref*{fig:qvh-phase_diagram}.
A charge density wave, illustrated in Fig. \subref*{fig:cdw}, develops and persists until the Hubbard constant reaches a critical value at which an antiferromagnetic ordering, depicted in Fig. \subref*{fig:af}, occurs.
For $\tilde{V}=0$, this critical value is $\tilde{U}_c \sim 2t$, as expected for honeycomb systems.\cite{PhysRevLett.111.036601}

From a mean-field perspective, the charge density wave groundstate leads to a change in the sublattice imbalance.
In other words, it leads to a transformation $m \to m + m_{\mathrm{CDW}}$.
The effects of an antiferromagnetic lead to the additional term in the Hamiltonian \ref{eq:topo-effetive_tb}:
\begin{align}\label{eq:af_mass}
  H_{AF} = m_{\mathrm{AF}} \kappa_0 \otimes \tau_0 \otimes \sigma_z \otimes s_z,
\end{align}
where $s_z$ acts on the spin space.
Note that one can interpret the antiferromagnetic term as a spin-dependent sublattice imbalance.

As discussed in Sec. \ref{sec:qvh-system}, small variations on the ratio $m/M$ lead to a gap opening.
That makes one wonder if there are topological phase transitions as we change the electron-electron coupling constants.
Hence, we compute, for different values of $m$ and $m_{\mathrm{AF}}$, the valley Chern number\cite{PhysRevB.84.205137, wolfElectricallyTunableFlat2019}
\begin{align}\label{eq:qvh-chern}
  &C_V = C_K - C_{K'} \nonumber \\
  &=\int_{-\infty}^0 d \omega
\int_{\scriptscriptstyle\textrm{BZ}} \frac{d^2 \mathbf k}{(2\pi)^2} \frac{\epsilon_{\alpha \beta}}{2}
 G_V (\partial_{k_\alpha}G_V^{-1}) (\partial_{k_\beta}G_V).
\end{align}
Here, $\epsilon_{\alpha\beta}$ denotes  the Levi-Civita tensor,
\begin{align}
G_V = [\omega - H(\mathbf k)+i0^+]^{-1}
\mathcal{P}_V
\end{align}
the Green's function associated with the Bloch Hamiltonian $H(\mathbf k)$, and $\mathcal{P}_V = \tau_z$ is the valley operator.

\begin{figure}[t!]
    \includegraphics[width=0.8\linewidth]{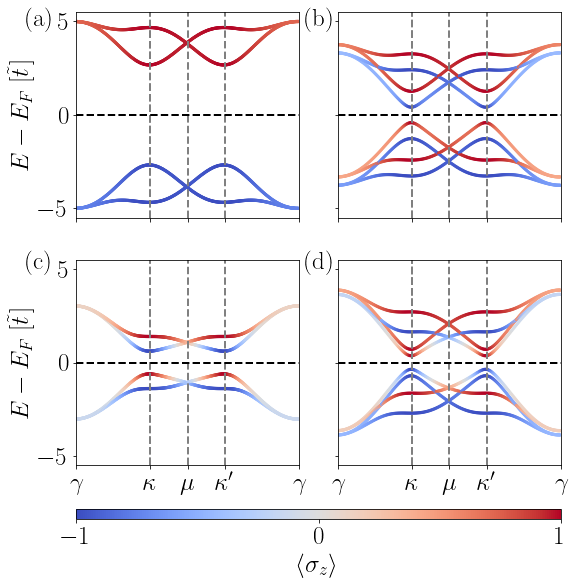}
  \caption{
  Sublattice-projected bandstructure for selected $\tilde{U}$ and $\tilde{V}$ showing all four distinct phases: (a) charge density wave, (b) antiferromagnet, (c) quantum valley Hall insulator, and (d) spin-polarized quantum valley Hall insulator.
  }
  \label{fig:various_bands}
\end{figure}

We see in Fig. \ref{fig:qvh-phases}a that there are two topologically nontrivial phases: one of them is a QVHI for both spin channels (valley Chern number 4), the other is the one that we predict to exist in a single spin channel (valley Chern number 2).
The formation of both phases is rather intuitive to understand.
While the QVHI takes place when $\delta(m / M) < 0$, the sQVHI phase exist for both positive and negative values of $m_{\mathrm{AF}}$.
The reason is that while one spin channel suffers a shift $m \to m + m_{\mathrm{AF}}$, the other is shifted as $m \to m - m_{\mathrm{AF}}$.
Thus, one spin channel becomes topological and the other becomes a trivial insulator.
This is visible in Fig. \ref{fig:various_bands}: in the QVHI phase [Fig. \ref{fig:various_bands} (c)], all four bands show a band inversion; and in the sQVHI phase [Fig. \ref{fig:various_bands} (d)], only two out of the four bands show a band inversion.
The band inversion occurs in the spin channels for which the sublattice imbalance decreases.

\begin{figure}
  \subfloat[\label{fig:valley_periodic}]{\includegraphics[height=.16\textheight]{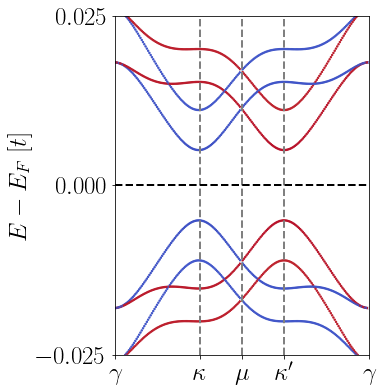}}
  \subfloat[\label{fig:valley_exp_qvh}]{\includegraphics[height=.175\textheight]{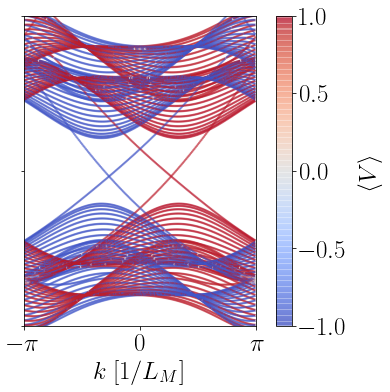}}\\
  \vspace*{-0.5cm}
  \subfloat[\label{fig:magn}]{\includegraphics[height=.145\textheight]{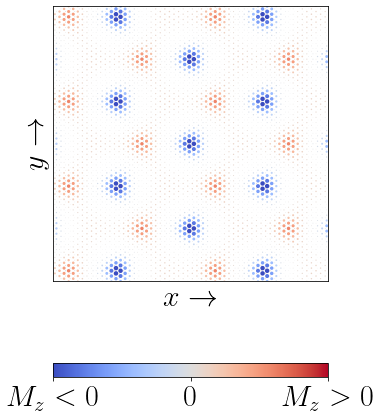}}
  \subfloat[\label{fig:y_exp_qvh}]{\includegraphics[height=.175\textheight]{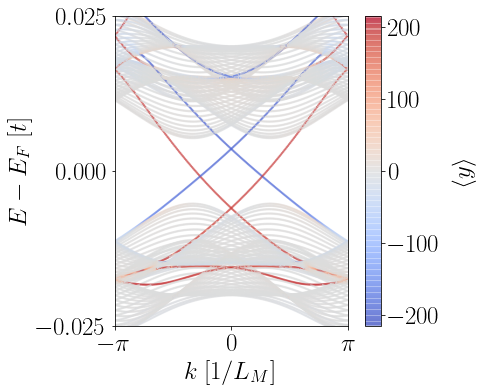}}
  \caption{
    Results of full-lattice calculations with modulated hoppings in Eq. \ref{eq:hoppings} solved self-consistently with an additional Hubbard constant $U=0.3t$.
    Valley Chern number in a (a) infinite system and (b) in a nanoribbon.
    (c) Magnetization along the $z$-direction.
    A periodically modulated ferrimagnetic order is formed.
    From a superlattice perspective, it cooresponds to a antiferromagnetic honeycomb structure.
    (d) $y$-position projection of a nanoribbon bandstructure. 
    We see that each edge has two counter-propagating edge states with opposite valley-polarization.
    }
  \label{fig:qvhi}
\end{figure}

Now we note that, even though there are only two regions in the order parameter map of Fig. \ref{fig:qvh-phase_diagram}, there are several gap closings in Fig. \ref{fig:qvh-gap}.
Reviewing the phase diagram by also checking the valley Chern number, it is visible that varying $\tilde{U}$ and $\tilde{V}$ yields four distinct phases: trivial charge density wave and antiferromagnet, as well as topological charge density wave (QVHI) and topological antiferromagnet (sQVHI).
We also show selected sublattice-projected band diagrams in Figs. \ref{fig:various_bands} a-d.
While the charge density wave (Fig. \ref{fig:various_bands}a) and antiferromagnetic (Fig. \ref{fig:qvhi}b) bulk bands have well-defined sublattice numbers, the QVHI (Fig. \ref{fig:various_bands}c) and the sQVHI (Fig. \ref{fig:various_bands}d) show a band inversion.

To compare with the effective model calculations, we perform self-consistent Hubbard mean-field calculations in a superlattice following our previous work.\cite{manescoCorrelationsElasticLandau2020}
To reduce the computational cost, we rescaled the system as $a \to \beta a$ and $t \to t / \beta $ so the Fermi velocity keeps constant. \cite{liuScalableTightBindingModel2015}
We choose, as an example, the Hubbard constant to be $U = 0.3t$, resulting in a correlation gap in the bandstructure (see Fig. \ref{fig:valley_periodic}).
The gap is  a result of a magnetic phase in the form of a superlattice-modulated ferrimagnetic order (see Fig. \ref{fig:magn}).
Integrating the magnetization in the vincinity to a minimum of $B(\vec{r})$ the magnetization is finite and positive, while it is negative in the neighborhood of pseudo-magnetic field maxima.
In this situation, the system has a valley Chern number 2, in accordance with our effective model calculations with an antiferromagnetic ordering.
We found that this system is in the sQVHI phase, with valley Chern number 2, compatible with the magnetically ordered groundstate observed.\cite{manescoCorrelationsElasticLandau2020} The existence of topological edge states is visible in the  bandstructure of a nanoribbon, shown in Fig. \ref{fig:valley_exp_qvh} and \ref{fig:valley_exp_qvh}.
As expected, we observe two counter-propagating (helical-like) edge states with opposite valley numbers at both boundaries, similarly to the Kane-Mele model.\cite{kaneQuantumSpinHall2005, kaneTopologicalOrderQuantum2005}
Finally, we note that the local charge accumulation with the periodic potential might also change the values of $\tilde{U}$ and $\tilde{V}$.
Furthermore, the increase of out-of-plane fields closes the antiferromagnetic gap.\cite{manescoCorrelationsElasticLandau2020}
Thus, electrostatic control is not only a knob to control topology, but also electronic correlations.

\section{Conclusions}
\label{sec:qvh-conclusion}
To summarize, we have shown that buckled graphene superlattices show spontaneous symmetry breaking driven by electronic interactions, leading to a topological gap opening.
First, by combining atomistic low energy models with a symmetry analysis, we derived an effective model for the lowest bands of buckled graphene superlattices.
We then included electronic interactions in a non-local form in the low-energy model,
showing the emergence of competing ground states.
Namely, an antiferromagnetic and a charge density wave.
Remarkably, the spontaneous breaking of symmetries was shown to lead to a topological gap opening for a wide range of the non-local interactions.
The charge density wave phase hosts a quantum valley Hall insulator, 
while the antiferromagnetic phase has a spin-polarized quantum valley Hall insulator region in the parameter space.
Our results put forward buckled graphene superlattices as a platform to study
interaction-induced valley topology, and highlight that single layer moire systems
can potentially host analogous correlated states to those of complex twisted graphene
multilayers.

\emph{Data availability}
The data shown in the figures, as well as the code generating all of the data is available on Zenodo.\cite{antonio_l_r_manesco_2021_4685133}

\emph{Acknowledgements.}
We thank Gabrielle Weber, Anton Akhmerov, and Eva Andrei for useful discussions.
The work of A.L.R.M. was funded by São Paulo 
Research Foundation, numbers 2016/10167-8 and 2019/07082-9.
J.L.L. acknowledges the computational resources provided by the Aalto Science-IT project
and the
financial support from the
Academy of Finland Projects No.
331342 and No. 336243.
A.L.R.M. also acknowledges the hospitality of the Quantum Tinkerer group.

\bibliographystyle{apsrev4-1}
\bibliography{refs}{}

\end{document}